\documentclass[twocolumn,showpacs,preprintnumbers,amsmath,amssymb]{revtex4}

\usepackage{bm}
\usepackage[colorlinks=true,linkcolor=blue,citecolor=blue,urlcolor=blue]{hyperref}
\usepackage{color}

\usepackage{graphicx}
\usepackage{epstopdf}

\begin{document}


\title{Large difference between the magnetic properties of Ba and Ti co-doped BiFeO$_3$ bulk materials and their corresponding nanoparticles prepared by ultrasonication} 


\author{Bashir Ahmmad}
\email[Author to whom correspondence should be addressed (e-mail): ]{arima@yz.yamagata-u.ac.jp}
\affiliation{Graduate School of Science and Engineering, Yamagata University, 4-3-16 Jonan, Yonezawa 992-8510, Japan.}

\author{Kensaku Kanomata, Kunihiro Koike,  Shigeru Kubota, Hiroaki Kato, Fumihiko Hirose}
\affiliation{Graduate School of Science and Engineering, Yamagata University, 4-3-16 Jonan, Yonezawa 992-8510, Japan.}

\author{Areef Billah, M. A. Jalil}
\author{M. A. Basith}
\email[Author to whom correspondence should be addressed (e-mail): ]{mabasith@phy.buet.ac.bd}
\affiliation{Department of Physics, Bangladesh University of Engineering and Technology, Dhaka-1000, Bangladesh.
}

\date{\today}

\begin{abstract}
	The ceramic pellets of the nominal compositions Bi$_{0.7}$Ba$_{0.3}$Fe$_{1-x}$Ti$_x$O$_3$ (x = 0.00-0.20) were prepared initially by standard solid state reaction technique. The pellets were then ground into micrometer-sized powders and mixed with isopropanol in an ultrasonic bath to prepare nanoparticles. The X-ray diffraction patterns demonstrate the presence of a significant number of impurity phases in bulk powder materials. Interestingly, these secondary phases were completely removed due to the sonication of these bulk powder materials for 60 minutes. The field and temperature dependent magnetization measurements exhibited significant difference between the magnetic properties of the bulk materials and their corresponding nanoparticles. We anticipate that the large difference in the magnetic behavior may be associated with the presence and absence of secondary impurity phases in the bulk materials and their corresponding nanoparticles, respectively. The leakage current density of the bulk materials was also found to suppress in the ultrasonically prepared nanoparticles compared to that of bulk counterparts.
	
   
\end{abstract}

\maketitle
\section{Introduction} \label{I}

Multiferroic materials, in which spontaneous ferroelectric polarization and magnetic order coexist, have been investigated intensively due to their multifunctionality \cite{ref1,ref2,ref3,ref4,ref5,ref6} and potentials for applications in the next-generation multifunctional devices. The co-existence of 'ferro'-orders in multiferroics allows the possibility that the magnetization can be tuned by an applied electric field and vice versa. Among compounds that are multiferroics, BiFeO$_3$ (BFO) is a paradigmatic and currently the most studied material. The multiferroic BFO exhibits the co-existence of ferroelectric ordering with Curie temperature (T$_C$) of 1103 K and antiferromagnetic (AFM) ordering with a N\'eel temperature (T$_N)$ of 643 K \cite{ref7}. However, preparation of pure BiFeO$_3$ is a challenge due to the formation of different impurity phases \cite{ref78, ref79}. The ferroelectricity in BiFeO$_3$ is considered to originate primarily from displacements of the Bi$^{3+}$ ions due to the lone 6S${^2}$ pair. The magnetic ordering of BiFeO$_3$ is G-type antiferromagnetic, having a spiral modulated spin structure (SMSS) with an incommensurate long-wavelength period of 62 nm \cite{ref7}. This spiral spin structure cancels the macroscopic magnetization and prevents the observation of the linear magnetoelectric effect \cite{ref8}. These problems ultimately limit the use of bulk BiFeO$_3$ in functional applications.  Many current investigations seek to suppress the spiral spin structure in an effort to release the inherent magnetization of this canted antiferromagnet and consequently to improve its multiferroic properties. One of the easiest ways to destroy the SMSS in BFO is the structural modifications or deformations introduced by cation substitutions or doping. Therefore, to perturb the SMSS and to improve the multiferroic properties of BiFeO$_3$, investigations were carried out substituting Bi by rare-earth ions \cite{ref9,ref10}  or alkaline-earth ions \cite{ref11}  and also substituting Fe by transition metal ions \cite{ref12, ref13}. Recent investigations also demonstrated that co-doping at Bi- and Fe-sites of BiFeO$_3$ by ions such as La and Mn \cite{ref14}, La and Ti \cite{ref15}, Nd and Sc \cite{ref16}, Gd and Ti \cite{ref17}, Ba and Mn \cite{ref18} etc., respectively can significantly improve multiferroic properties of BiFeO$_3$ for various applications. In this investigation, we have preferred this co-doping approach and have performed simultaneous minor substitution of Bi and Fe in BiFeO$_3$ by ions such as Ba and Ti, respectively to observe their multiferroic properties.
The partial substitution of Bi$^{3+}$ with ions having the biggest ionic radius can suppress effectively the spiral spin structure of BiFeO$_3$ \cite{ref19}, therefore, we have chosen Ba (ionic radii 1.42 \AA) in place of Bi (ionic radii 1.17 \AA) as a doping element. Moreover, it is already proved that substitution of Ba in place of Bi in BiFeO$_3$ can improve their multiferroic properties \cite{ref20}. The substitution of Ti in stead of Fe in BiFeO$_3$ is especially attractive due to the fact that it is able to decrease the leakage current in BiFeO$_3$ \cite{ref21}. Therefore, we think that co-doping of Ba and Ti in BiFeO$_3$ could be an effective route to improve magnetic as well as ferroelectric properties of BiFeO$_3$. 

The spiral spin structure of BFO is also known to be destroyed by the action of finite size effects in nanophase BiFeO$_3$ \cite{ref22}.  It is evident that nanoparticles of BiFeO$_3$, especially those with a particle size of the order of or smaller than the 62 nm SMSS, exhibit improved multiferroic properties \cite{ref22, ref23,ref24}. Synthesis of BiFeO$_3$ multiferroic nanoparticles hence requires a particle size of the order of or smaller than 62 nm. The most widely used synthesis techniques for BiFeO$_3$ nanoparticles are based on different chemical routes \cite{ref22, ref25,ref26} which involve toxic precursors and complex solution processes \cite{ref27}. Recently, we have developed a physical technique using ultrasonic energy to produce high quality monodisperse Gd and Ti co-doped BiFeO$_3$ nanoparticles \cite{ref23} directly from their bulk powder materials. In this present investigation, this ultrasonication technique was used to prepare Bi$_{0.7}$Ba$_{0.3}$Fe$_{1-x}$Ti$_x$O$_3$ (x = 0.00-0.20) nanoparticles directly from their bulk powder materials. 

It should be noted that some investigations were already carried out on Ba and Ti co-doped BiFeO$_3$ bulk multiferroic material system \cite{ref28, ref29}, however, a systematic investigation on the preparation and characterization of their nanoparticles and comparison of their properties with bulk materials is still unexplored to the best of our knowledge. Therefore, we have prepared Ba and Ti co-doped BiFeO$_3$ nanoparticles by ultrasonication of their bulk powder materials and compared the magnetic and ferroelectric properties between nanoparticles and bulk materials.

\section{Experimental details} \label{II}
The polycrystalline samples of nominal compositions Bi$_{0.7}$Ba$_{0.3}$Fe$_{1-x}$Ti$_x$O$_3$ (x = 0.00-0.20) were synthesized by using standard solid state reaction technique as was described in details in our previous investigation \cite{ref17} for Gd and Ti co-doped Bi$_{0.9}$Gd$_{0.1}$Fe$_{1-x}$Ti$_x$O$_3$ (x = 0.00-0.25) material system. The analytical grade oxides of Bi$_{2}$O$_{3}$, Ba$_{2}$O$_{3}$, Fe$_{2}$O$_{3}$, and TiO$_{2}$ powders were carefully weighed in stoichiometric proportion, mixed thoroughly with acetone and  ground in an agate mortar until a homogeneous mixture was formed. The compacted mixtures of reagents taken in desired cation ratios were calcined at 750$^o$C for 2 hours in a programmable furnace. The calcined powders were ground again for 2 hours to get more homogeneous mixture. The powders were pressed into pellets of thickness 1 mm and diameter 12 mm by using a uniaxial hydraulic press and sintered at 850$^o$C for 5 hours at heating rate 3$^o$C per minute. The powder materials and ceramic pellets were used for structural, dielectric and magnetic measurements of bulk materials. 
\begin{figure}[hh]
	\centering
	\includegraphics[width=8.5cm]{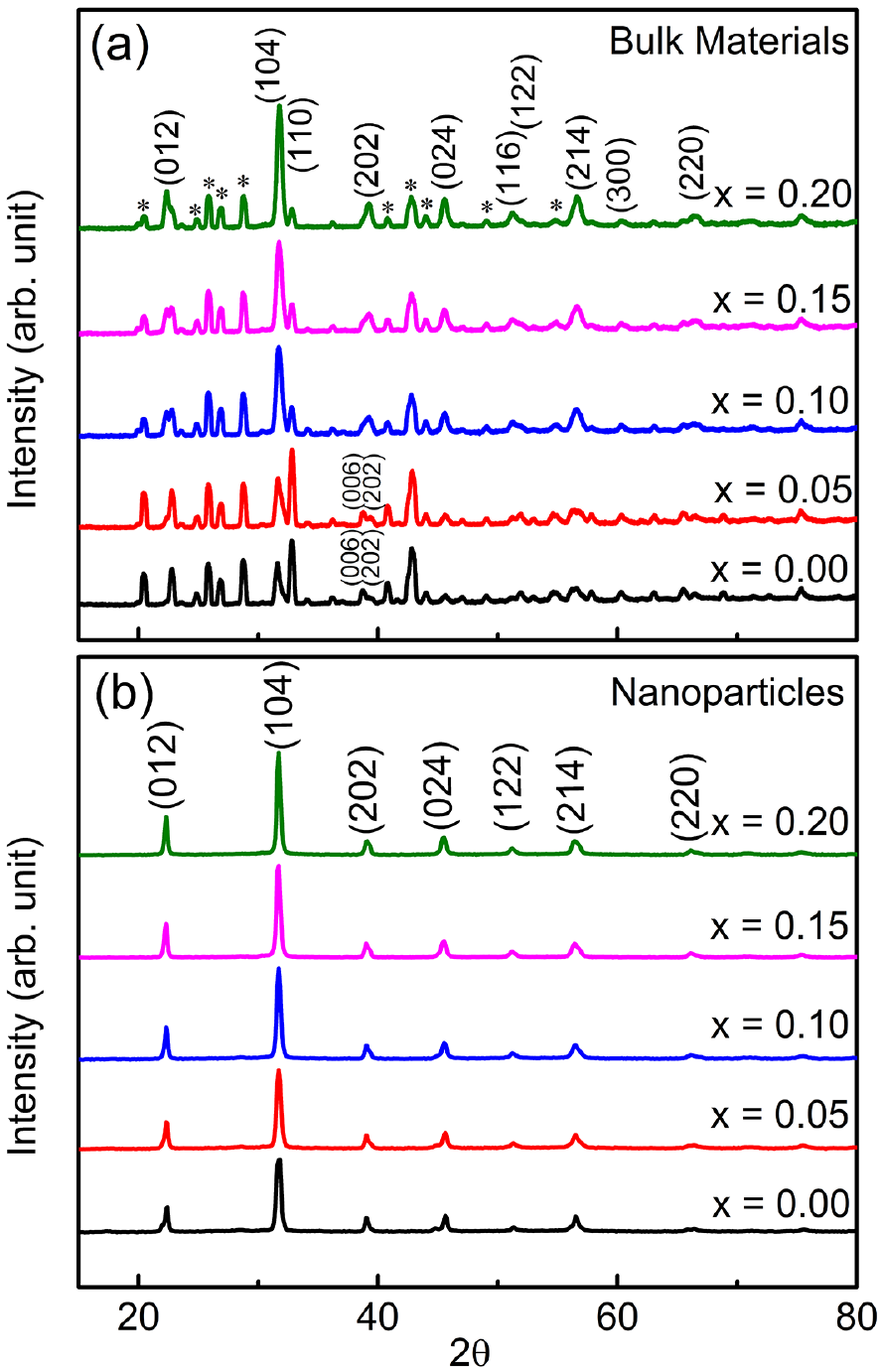}
	\caption{(a) X-ray diffraction patterns of the Bi$_{0.7}$Ba$_{0.3}$Fe$_{1-x}$Ti$_{x}$O$_3$ (x = 0.00, 0.05, 0.10, 0.15 and 0.20) bulk samples and their (b)  nanoparticles. } \label{fig1}
\end{figure}

The ceramic pellets  were  ground again  into powder by manual grinding.  The Bi$_{0.7}$Ba$_{0.3}$Fe$_{1-x}$Ti$_x$O$_3$ (x = 0.00-0.20) nanoparticles were prepared directly from this bulk powder by using the sonication technique described in Ref. \cite{ref23}. The bulk powders were mixed with isopropanol with a ratio of 1 g powder and 100 ml isopropanol and then put into an ultrasonic bath and was sonicated for 60 minutes. After around four hours, $\sim$ 40 \% of the mass was collected as supernatant and was used for structural and magnetic characterization. To measure electrical properties of the synthesized nanoparticles, pellets were prepared by pressing and sintering at 850$^o$C with high heating rate i.e. 20$^o$C/min \cite{ref24}.


The crystal structure of the nanoparticles as well as bulk powder materials were determined from X-ray diffraction (XRD) data using a diffractometer (Rigaku SmartLab) with CuK$_{\alpha}$  (${\lambda}$ = 1.5418 $\AA$) radiation. The size of the synthesized nanoparticles was studied using transmission electron microscopy (TEM) imaging. The $M-H$ hysteresis loops of Bi$_{0.7}$Ba$_{0.3}$Fe$_{1-x}$Ti$_x$O$_3$ (x = 0.00-0.20) multiferroic ceramics and their nanoparticles were carried out at room temperature using a vibrating sample magnetometer (VSM, Riken Denshi Co. Ltd.). The temperature dependent magnetization measurements were carried out both at zero field cooling (ZFC) and field cooling (FC) processes \cite{ref30} using a Superconducting Quantum Interference Device (SQUID) Magnetometer (Quantum Design MPMS-XL7, USA). X-ray photoelectron spectroscopy (XPS, ULVAC-PHI Inc., Model 1600) analysis was carried out with a Mg-K{$\alpha$} radiation source to know oxygen concentration. The leakage current density and ferroelectric polarization of the pellet shaped samples of bulk and nanoparticles were carried out using a ferroelectric loop tracer in conjunction with external amplifier (10 kV) (Marine India).


\section{Results and discussions} \label{III}
\subsection{Structural characterization} \label{I}
The structural properties of Bi$_{0.7}$Ba$_{0.3}$Fe$_{1-x}$Ti$_{x}$O$_3$ (x = 0.00-0.20) ceramics and their corresponding nanoparticles  were analyzed with powder XRD. The XRD patterns of the Bi$_{0.7}$Ba$_{0.3}$Fe$_{1-x}$Ti$_{x}$O$_3$ (x = 0.00-0.20) bulk materials and their corresponding nanoparticles with rhombohedral crystal structure are shown in figures \ref{fig1}(a) and (b), respectively. In the bulk materials prepared by solid state reaction technique, a significant number of secondary phases appeared as shown in figure \ref{fig1}(a). These secondary phases may be a Bi-rich Bi$_{25}$FeO$_{39}$ (ICDD File Card No. 77-0865), Bi$_2$O$_3$ (ICDD card No.76-2478) and an Fe-rich
Bi$_2$Fe$_4$O$_9$ phase (ICDD File Card No. 25-0090) \cite{ref27, ref107}. During the solid sate synthesis of bulk undoped BiFeO$_3$ and cations substituted BiFeO$_3$ the formation of these secondary phases were observed in a number of previous investigations \cite{ref27, ref107, ref108, ref109}. In a previous investigation, it was observed that the specific reaction pathway has good influence on the final distribution of the secondary phases \cite{ref38}, however, the mechanism leading to the formation of those secondary phases is still unknown. Interestingly, these secondary phases were found to remove completely in ultrasonically \cite{ref23} prepared nanoparticles \cite{ref30}, figure \ref{fig1}(b). 

In the XRD patterns, figure \ref{fig1}(a), the (110) and (006) peaks are associated with twin peaks \cite{ref121} of bulk materials. For x = 0.00 and 0.05 compositions, the (110) (2$\theta$ = 33$^{\circ}$) peak was intense compared than that of (104) (2$\theta$ = 32$^{\circ}$) peak, whereas for x = 0.10-0.20 compositions the opposite trend of this intensity value was observed. The (006) (2$\theta$ = 38.5$^{\circ}$) peak is appeared for x = 0.00 and 0.05 compositions, however, for a further increment of Ti concentration, the (006) peak was eliminated. In another investigation, due to substitution of Mn in place of Fe in BiFeO$_3$, the (006) peak was also appeared at the low concentration of Mn, whereas for a larger concentration of Mn this peak was eliminated \cite{ref122}. In the case of ultrasonically prepared nanoparticles along with impurity peaks the (110) and (006) peaks were also eliminated.

\begin{figure}[hh] 
	\centering
	\includegraphics[width=8.5cm]{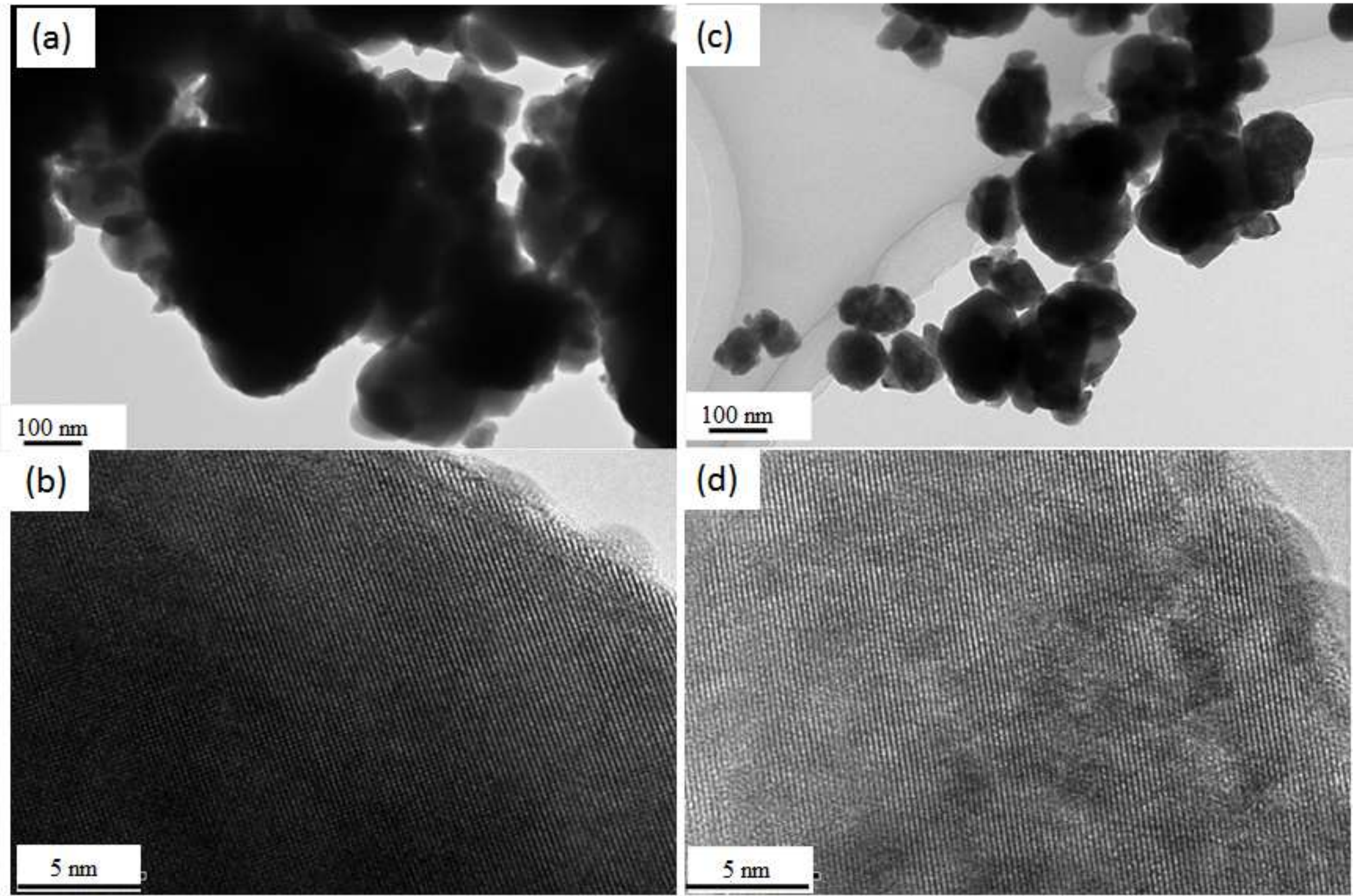}
	\caption{(a) The bright field TEM image demonstrates the particles of Bi$_{0.7}$Ba$_{0.3}$Fe$_{0.9}$Ti$_{0.1}$O$_3$  bulk materials. (b) HRTEM image showing the crystal plane of a particle. (c) The BF image shows the size of Bi$_{0.7}$Ba$_{0.3}$Fe$_{0.9}$Ti$_{0.1}$O$_3$ particles obtained for 60 minutes sonication of the bulk powder materials. (d) HRTEM image showing the crystallinity of the ultrasonically prepared nanoparticles.} \label{fig2}
\end{figure}

The full width at half maximum intensity (FWHM) of the most intense diffraction peak (104) of figure \ref{fig1}(b) was utilized in Scherrer equation to calculate average crystallite sizes of the ultrasonically prepared samples. The calculated average crystallite sizes of ultrasonically prepared Bi$_{0.7}$Ba$_{0.3}$Fe$_{1-x}$Ti$_{x}$O$_3$ samples are 19 nm, 20 nm, 22 nm, 23 nm and 24 nm for x = 0.00, 0.05, 0.10, 0.15 and 0.20, respectively. The bulk morphology of the representative Bi$_{0.7}$Ba$_{0.3}$Fe$_{0.9}$Ti$_{0.1}$O$_3$ ceramics prepared by solid state reaction technique is shown in figure \ref{fig2}(a). Due to the high degree of agglomeration of the particles as was seen in TEM bright field image, it was really difficult to estimate the particle size. However, the high resolution TEM image demonstrates the good crystallinity of the bulk materials, figure \ref{fig2} (b). Figure \ref{fig2} (c) shows bright-field TEM image of representative Bi$_{0.7}$Ba$_{0.3}$Fe$_{0.9}$Ti$_{0.1}$O$_3$ particles obtained after sonication for 60 minutes. Although the agglomeration of the synthesized nanoparticles is not severe like the bulk ceramic materials, however, it is still difficult to identify the exact size of the nanoparticles from TEM image. The sizes of the ultrasonically prepared Bi$_{0.7}$Ba$_{0.3}$Fe$_{0.9}$Ti$_{0.1}$O$_3$ particles roughly vary from 50 to 200 nm, which is larger than that calculated by the Scherrer equation. The large particle size determined by TEM images compared than that calculated by Scherrer equation has also been reported in previous investigations \cite{ref101,ref102} and this was due to the agglomeration of the particles.

Figure \ref{fig2} (d) demonstrates high resolution (HR) TEM image of Bi$_{0.7}$Ba$_{0.3}$Fe$_{0.9}$Ti$_{0.1}$O$_3$ nanoparticles obtained at 60 minutes of sonication. This HRTEM image is an evidence of the single crystal nanoparticles synthesized using a sonication time of 60 minutes. 

\subsection{Magnetic characterization} \label{II}
\begin{figure}[!hh]
	\centering
	\includegraphics[width=8.5cm]{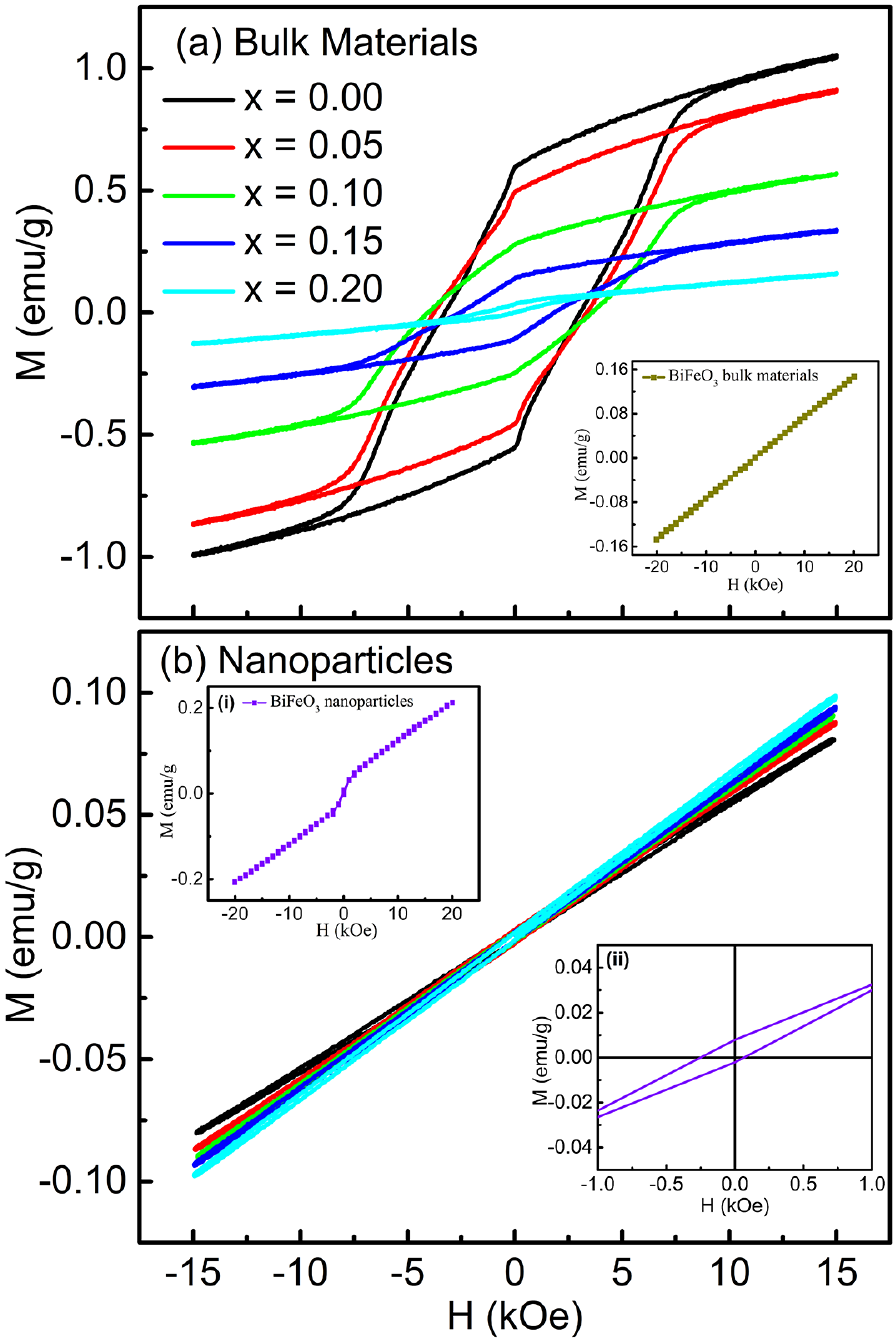}
	\caption{The room temperature $M-H$  hysteresis loops of Bi$_{0.7}$Ba$_{0.3}$Fe$_{1-x}$Ti$_{x}$O$_3$ (x = 0.00, 0.05, 0.10, 0.15 and 0.20) (a) bulk materials and (b) nanoparticles obtained after a sonication time of 60 minutes. The inset of figure (a) demonstrates the hysteresis loop of bulk undoped BiFeO$_3$. The inset of figure (b) (i) shows the hysteresis loop of undoped BiFeO$_3$ nanoparticles prepared by ultrasonication. The enlarged view of this hysteresis loop is shown in figure (b) (ii).} \label{fig3}
\end{figure}

For magnetic characterization, the M-H hysteresis loops of Bi$_{0.7}$Ba$_{0.3}$Fe$_{1-x}$Ti$_{x}$O$_3$ (x = 0.00-0.20) bulk materials and nanoparticles were measured at room temperature with an applied magnetic field of up to $\pm$15 kOe. The room temperature M-H loops of Bi$_{0.7}$Ba$_{0.3}$Fe$_{1-x}$Ti$_{x}$O$_3$ (x = 0.00-0.20) bulk samples and their corresponding nanoparticles for x = 0.00, 0.05, 0.10, 0.15 and 0.20 are displayed in figure \ref{fig3} (a) and (b), respectively. The inset of figure \ref{fig3} (a) demonstrates the hysteresis loop of bulk undoped BiFeO$_3$. The inset of figure \ref{fig3} (b) (i) shows the hysteresis loop of BiFeO$_3$ nanoparticles prepared by ultrasonication. The enlarged view of this hysteresis loop is shown in figure \ref{fig3} (b) (ii). The linear M-H curve of bulk undoped BiFeO$_3$ demonstrates its antiferromagnetic nature \cite{ref17}. Unlike undoped bulk BiFeO$_3$, the Ba doped Bi$_{0.7}$Ba$_{0.3}$FeO$_3$ as well as Ba and Ti co-doped bulk Bi$_{0.7}$Ba$_{0.3}$Fe$_{1-x}$Ti$_{x}$O$_3$ (x = 0.00-0.20) materials exhibit nearly ferromagnetic behavior with notable remanent magnetization and coercivity. The ferromagnetic behavior of Ba and Ti co-doped bulk BiFeO$_3$ multiferroics is consistent with previous investigations \cite{ref28, ref29}.  

\begin{figure}[!hh]
	\centering
	\includegraphics[width=7.2cm]{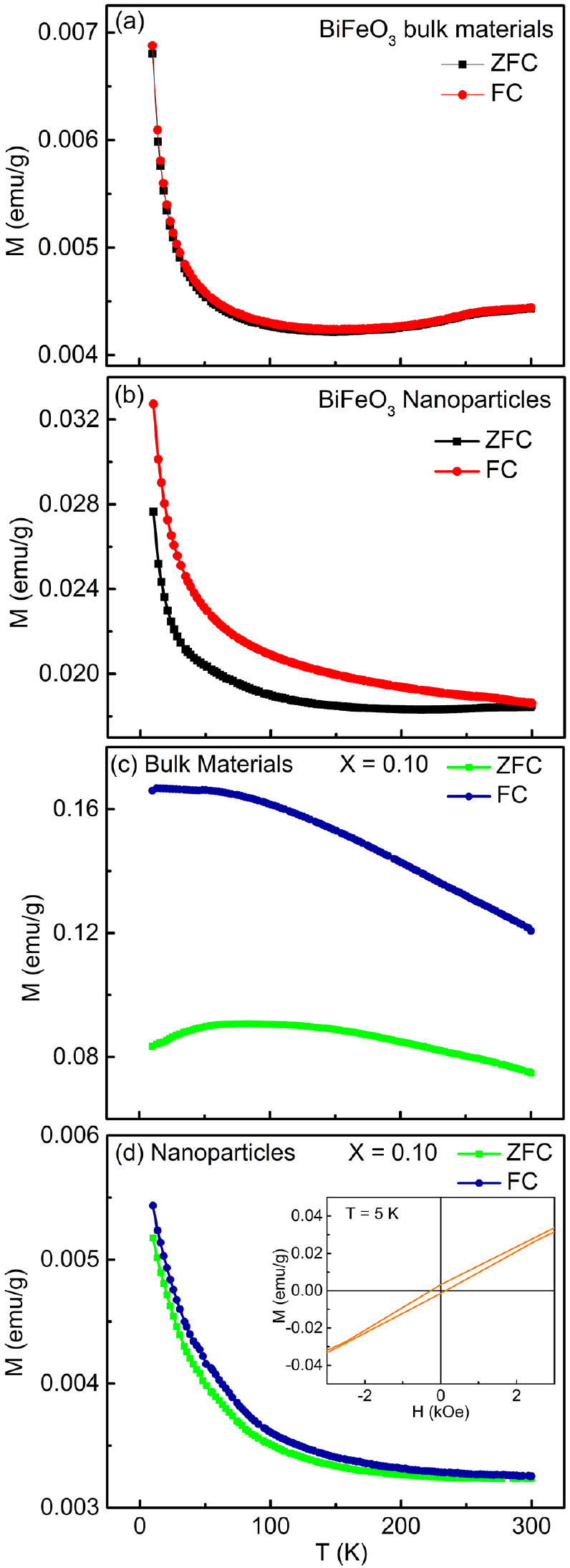}
	\caption{Temperature dependent ZFC and FC magnetization curves of undoped BiFeO$_3$ (a) bulk materials and (b) corresponding nanoparticles sonicated for 60 minutes. Temperature dependence of magnetization of Bi$_{0.7}$Ba$_{0.3}$Fe$_{0.9}$Ti$_{0.1}$O$_3$ bulk materials and their corresponding nanoparticles obtained after a sonication time of 60 minutes are shown in figures (c) and (d), respectively.} \label{fig6}
\end{figure}

From the hysteresis loops, the remanent magnetization (M$_r$) and coercive fields (H$_c$) were quantified as: M$_{r}$ = $|$(M$_{r1}$-M$_{r2}$)$|$/2 where M$_{r1}$ and M$_{r2}$ are the magnetization with positive and negative points of intersection with H = 0, respectively \cite{ref17} and $H_c = (H_{c1}-H_{c2})/2$, where H$_{c1}$ and H$_{c2}$ are the left and right coercive fields \cite{ref17, ref30}. Calculated values of M$_{r}$ and H$_c$ for x = 0.00, 0.05, 0.10, 0.15 and 0.20 bulk materials are inserted in table \ref{Tab1}. The remanent magnetization of bulk Bi$_{0.7}$Ba$_{0.3}$Fe$_{1-x}$Ti$_{x}$O$_3$ (x = 0.00-0.20) materials were found to decrease with increase of Ti concentration as was also observed in Ref. \cite{ref28} for similar multiferroic material system. On the other hand, the coercive fields were enhanced with Ti substitution of up to 10\% as was marked by red in table \ref{Tab1}. For a further increment of Ti concentration in place of Bi in BiFeO$_3$, the H$_c$ values were decreased. The trend of the variation of coercivity with Ti concentration is different to some extent than that reported in Ref. \cite{ref28} in which like M$_{r}$, H$_{c}$ values were also decreased with Ti substitution. The coercivity is indeed an extrinsic property and depends mostly on a number of factors like the microstructure, in particular the homogeneity of the grain, their size distribution, domain wall pinning effect ect and hence such a compositional variation of coercivity is not unexpected.

The room temperature $M$-$H$  hysteresis loops of Bi$_{0.7}$Ba$_{0.3}$Fe$_{1-x}$Ti$_{x}$O$_3$ (x = 0.00-0.20) bulk materials exhibit an asymmetric shift towards the magnetic field axes (figure \ref{fig3} (a)) \cite{ref17}. This indicates the presence of an exchange bias effect in this bulk material system \cite{ref31}. The exchange bias field (H$_{EB}$) can be quantified as $H_{EB} = -(H_{c1}+H_{c2})/2$ where $H_{c1}$ and $H_{c2}$ are the left and right coercive fields, respectively \cite{ref31}. The H$_{EB}$ values of Bi$_{0.7}$Ba$_{0.3}$Fe$_{1-x}$Ti$_{x}$O$_3$ (x = 0.00-0.20) bulk materials calculated from the asymmetric shift of the $M$-$H$ hysteresis loops of figure \ref{fig3}(a) is also inserted in table \ref{Tab1}. As shown in figure \ref{fig3} (a), the hysteresis loops of Bi$_{0.7}$Ba$_{0.3}$Fe$_{1-x}$Ti$_{x}$O$_3$ (x = 0.00-0.20) bulk materials is not completely saturated even with an applied magnetic field of up to $\pm$15 kOe which confirm the antiferromagnetic nature of the compounds. Moreover, compared to the undoped BiFeO$_3$ (inset of figure \ref{fig3} (a)) \cite{ref31}, the large values of M$_r$ and H$_c$ of bulk materials displayed in table \ref{Tab1} also indicate their ferromagnetic nature. Therefore, we may conclude the co-existence of ferri/ferromagnetic (FM) and antiferromagnetic domains in this multiferroic material system. As a consequence of the exchange coupling at interfaces between these multiple magnetic domains, it is expected that the system acts as a natural system for generating EB effect in Bi$_{0.7}$Ba$_{0.3}$Fe$_{1-x}$Ti$_{x}$O$_3$ multiferroics \cite{ref17, ref31, ref71, ref72}. The $H_{EB}$ values were found to enhance with Ti doping concentration (table \ref{Tab1}).

\begin{table}[!h]
	\caption{The table shows the remanent magnetization $M_{r}$, coercive fields $H_{c}$ and exchange bias fields $H_{EB}$ of Bi$_{0.7}$Ba$_{0.3}$Fe$_{1-x}$Ti$_x$O$_3$ (x = 0.00-0.20) bulk powder materials at room temperature.} \label{Tab1} 
	\begin{center}
		\begin{tabular}{|l|l|l|l|}
			\hline
			
			\cline{1-4}
			${x}$  (Ti &$M_{r}$ (emu/g)  &$H_{c}$ (kOe)  &$H_{EB}$ (kOe)\\
			${content)}$ & &&\\
			\hline
			0.00&$0.57$&$\textcolor{red}{3.15}$&$0.12$\\
			\hline
			0.05&$0.48$&$\textcolor{red}{3.55}$&$0.22$\\
			\hline
			0.10&$0.26$&$\textcolor{red}{3.81}$&$0.22$\\
			\hline
			0.15&$0.12$&$2.20$&$0.37$\\
			\hline
			0.20&$0.02$&$0.74$&$0.73$\\
			\hline
			\hline
		\end{tabular}
	\end{center}
\end{table}

Surprisingly, in the case of nanoparticles, figure \ref{fig3} (b), we have observed a completely different magnetic behavior than that of bulk materials. The hysteresis loops of the synthesized nanoparticles is completely unsaturated with negligible coercivity, figure \ref{fig3} (b). The linear M-H curves of the synthesized nanoparticles reveal their antiferromagnetic nature rather than the nearly ferromagnetic characteristic of their bulk counterparts. To a first consideration we thought this is a size effect of the nanoparticles and below a critical size, the particles are behaving like superparamagnetic nanoparticles. However, looking at bright field TEM image, figure \ref{fig2} (c), we can not be convinced that the particles are superparamagnetic in nature. It should be noted that the M-H hysteresis loop of undoped BiFeO$_3$ nanoparticles exhibit a small loop at the center of the hysteresis with a coercivity of 159 Oe and remanent magnetization of 0.005 emu/g as shown in the inset of figure \ref{fig3} (b) (ii). The high value of magnetization of BiFeO$_3$ nanoparticles compared to that of bulk BiFeO$_3$ materials is also worth noting. The ultrasonically prepared BiFeO$_3$ nanoparticles also exhibit an exchange bias effect (inset of figure \ref{fig3} (b) (ii)) with a biasing field of 93 Oe at room temperature. Thus the magnetic properties of undoped BiFeO$_3$ nanoparticles is also different than that of bulk counterpart.

To further explore the large difference in the magnetic properties of the bulk and nanoparticles of undoped BiFeO$_3$ as well as Ba and Ti co-doped BiFeO$_3$, we have carried out temperature dependent ZFC and FC magnetization measurements under a magnetic field of 500 Oe. In ZFC condition, the sample was initially cooled from 300 K to the lowest achievable temperature and data were collected while heating in the presence of the 500 Oe applied field. On the other hand, in the FC mode, data values were collected while cooling in the presence of the magnetic field  \cite{ref31}. The $M-T$ curves measured in ZFC and FC modes in the presence of 500 Oe applied magnetic field are shown in figures \ref{fig6} (a) and (b) for undoped BiFeO$_3$ and their corresponding nanoparticles, respectively. Figures \ref{fig6} (c) and (d) demonstrated the ZFC and FC curves of representative Bi$_{0.7}$Ba$_{0.3}$Fe$_{0.9}$Ti$_{0.1}$O$_3$ bulk materials and ultrasonically prepared corresponding nanoparticles, respectively.

The ZFC and FC curves of undoped bulk BiFeO$_3$ ceramics show an anomaly near 264 K (figure \ref{fig6} (a)). Similar anomaly was also observed in previous investigations and was disappeared by cation substitutions in BiFeO$_3$ ceramics \cite{ref34, ref103}. In Ref. \cite{ref103} it was anticipated that this anomaly  originates from domain wall pinning effects due to random distribution of oxygen vacancies. Interestingly, in the present investigation, this anomaly was found to disappear (figure \ref{fig6} (b)) due to the ultrasonication of the materials for 60 minutes. Notably, the magnetization in bulk BiFeO$_3$ decreases with the temperature from 300 K to 150 K suggesting the antiferromagnetic nature of the compounds up to 150 K below which especially below 50 K an abrupt increase of magnetization was observed. Moreover, in the bulk undoped BiFeO$_3$, both ZFC and FC magnetization curves overlap which also suggest their antiferromagnetic nature. In the case of BiFeO$_3$ nanoparticles, the magnetization value was gradually increased with decreasing temperature and yielded magnetization values larger than that measured for the bulk ceramics. Beside this, a small splitting between ZFC and FC curves  was observed which indicates the coexistence of AFM and FM orderings in BiFeO$_3$ nanoparticle system. This mixed magnetic ordering was also confirmed by an asymmetric shift of the M-H hysteresis loop (insets of figure \ref{fig3} (b) (ii)) of BiFeO$_3$ nanoparticles.




The temperature dependent magnetization curves,  figure \ref{fig6} (c), demonstrate the ferromagnetic nature of representative Bi$_{0.7}$Ba$_{0.3}$Fe$_{0.9}$Ti$_{0.1}$O$_3$ bulk materials \cite{ref35}. It is generally expected that due to Ba and Ti co-doping, the spin cycloid structure of BiFeO$_3$ can be destroyed, the latent magnetization locked within the cycloid structure is released and the samples become ferromagnetically ordered \cite{ref32}. The role of secondary phases appeared in the XRD patterns of bulk materials, figure \ref{fig1} (a) should also be considered and will be discussed later on. On the contrary, the ZFC and FC M-T curves of Bi$_{0.7}$Ba$_{0.3}$Fe$_{0.9}$Ti$_{0.1}$O$_3$ nanoparticles (figure \ref{fig6} (d) are completely different than that of bulk materials. The magnetization of nanoparticles increases slowly as temperature is decreasing to 150 K, and then sharply rises up with further decreasing temperature. The steep increase of magnetization particularly below 50 K in figures \ref{fig6} (a), (b) and (d) indicates the weak ferromagnetic nature of this material system at sufficiently low temperature \cite{ref30, ref31, ref34}. To further confirm the weak ferromagnetism at low temperature in the ultrasonically prepared Bi$_{0.7}$Ba$_{0.3}$Fe$_{0.9}$Ti$_{0.1}$O$_3$ nanoparticles, an M-H hysteresis loop was carried out at 5 K as shown in the inset of figure \ref{fig6} (d). A tiny loop at the center of the hysteresis with a coercivity of 196 Oe and remanent magnetization of 0.002 emu/g was observed at 5 K. Similar to the ultrasonically prepared undoped BiFeO$_3$ nanoparticles, at 5 K the Ba and Ti co-doped Bi$_{0.7}$Ba$_{0.3}$Fe$_{0.9}$Ti$_{0.1}$O$_3$ nanoparticles also show an exchange bias effect (inset of figure \ref{fig6} (d) with a biasing field of 69 Oe. This indicates the coexistence of AFM and FM orderings in Ba and Ti co-doped Bi$_{0.7}$Ba$_{0.3}$Fe$_{0.9}$Ti$_{0.1}$O$_3$ nanoparticle system at low temperature. It should also be noted clearly that in undoped BiFeO$_3$ nanoparticles as well as Ba and Ti co-doped Bi$_{0.7}$Ba$_{0.3}$Fe$_{1-x}$Ti$_{x}$O$_3$ bulk ceramics, the exchange bias effect was also observed at room temperature. On the contrary, in the case of Ba and Ti co-doped Bi$_{0.7}$Ba$_{0.3}$Fe$_{1-x}$Ti$_{x}$O$_3$ nanoparticle system, we didn't observe the sign of any exchange bias effect at room temperature. The linear M-H hysteresis loops and the absence of any exchange bias effect in  Bi$_{0.7}$Ba$_{0.3}$Fe$_{1-x}$Ti$_{x}$O$_3$ nanoparticles clearly indicate their antiferromagnetic nature at room temperature similar to the bulk undoped BiFeO$_3$ ceramics.



The reasons behind this large difference in the magnetization of Ba and Ti co-doped Bi$_{0.7}$Ba$_{0.3}$Fe$_{1-x}$Ti$_{x}$O$_3$ bulk and nanoparticle systems are open. The magnetization in the phase pure Bi$_{0.7}$Ba$_{0.3}$Fe$_{1-x}$Ti$_{x}$O$_3$ nanoparticles are even smaller than that of undoped bulk BiFeO$_3$ materials. It should be noted that in our previous investigation \cite{ref23}, the magnetization of ultrasonically prepared Gd and Ti co-doped BiFeO$_3$ nanoparticles were much larger compared to their bulk counterparts. It was stated that due to the large effective magnetic moment of Gd$^{3+}$ (8.0 Bohr Magnetron), the ferromagnetic coupling between Gd$^{3+}$ and Fe$^{3+}$ ions contributed \cite{ref104} significantly to enhance the magnetization in Gd and Ti co-doped BiFeO$_3$ nanoparticle system. In the present investigation, the substitution of nonmagnetic Ba and Ti in BiFeO$_3$ reduced the magnetization in phase pure Ba and Ti co-doped Bi$_{0.7}$Ba$_{0.3}$Fe$_{1-x}$Ti$_{x}$O$_3$ nanoparticles. The net value of magnetization of Ba and Ti co-doped  BiFeO$_3$ nanoparticles is smaller than that of undoped and co-doped bulk counterparts. It is worth mentioning that the XRD patterns demonstrate the presence of huge secondary phases in the bulk materials, which were totally absent in the case of synthesized nanoparticles. We expect that the undesirable magnetic impurity phases played a significant role to enhance  ferromagnetism in Ba and Ti co-doped Bi$_{0.7}$Ba$_{0.3}$Fe$_{1-x}$Ti$_{x}$O$_3$ bulk materials \cite{ref27, ref75}. In the absence of these secondary impurity phases, the nonmagnetic doping elements Ba and Ti in BiFeO$_3$ played role to reduce the magnetization of the ultrasonically prepared phase pure nanooparticles.

In the next stage of this investigation, the electric properties of this multiferroic material system were investigated.



\subsection{Electric measurements} \label{IIIB}
\begin{figure}[!hh]
	\centering
	\includegraphics[width=8.5cm]{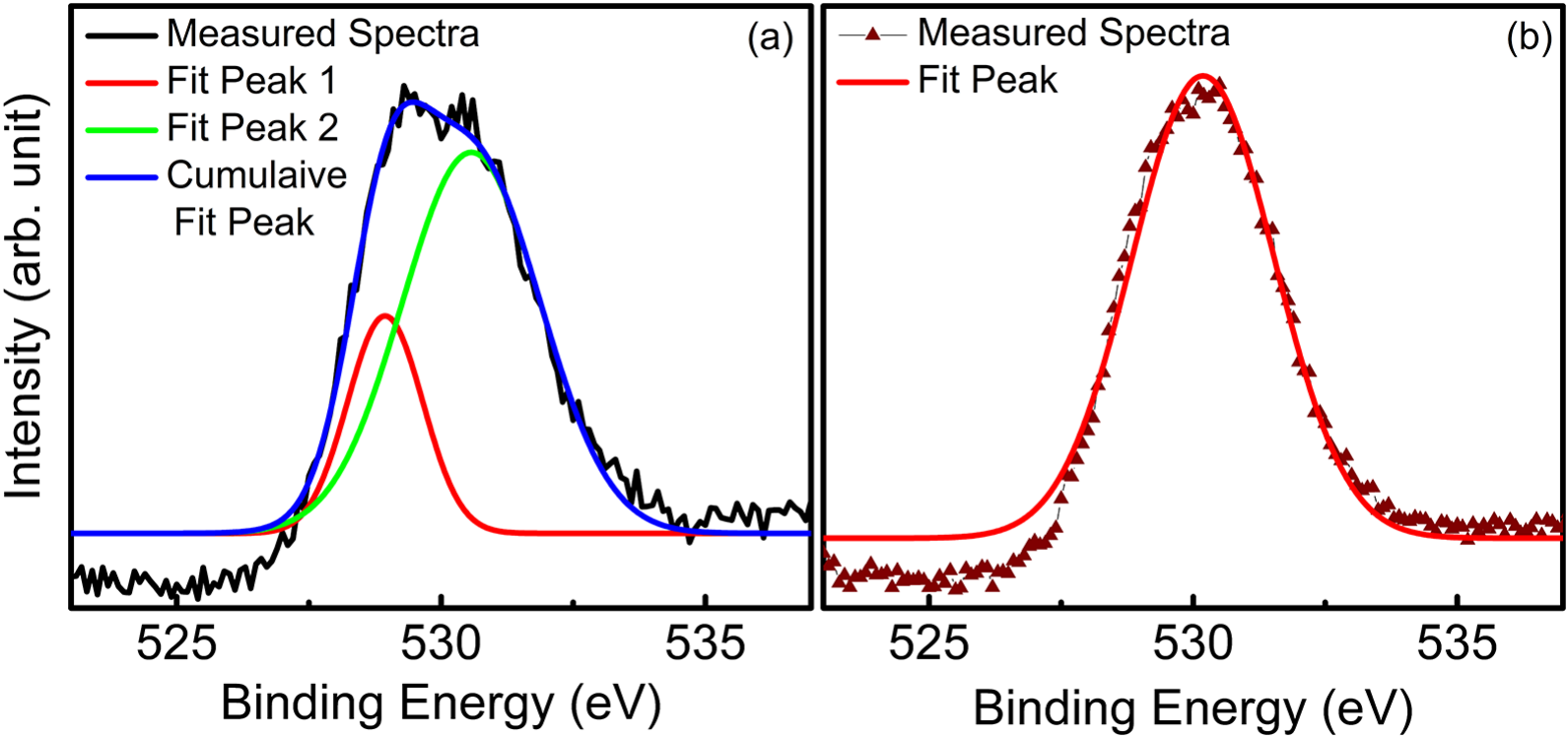}
	\caption{XPS spectra of the O 1s of Bi$_{0.7}$Ba$_{0.3}$Fe$_{0.9}$Ti$_{0.1}$O$_3$ (a) bulk polycrystalline powder materials sintered at $\mathrm{850^{0}C}$ and (b) nanoparticles sonicated for 60 minutes. } \label{fig7}
\end{figure}
Figure \ref{fig7}(a) displays the core level XPS spectra of O 1s orbital of representative Bi$_{0.7}$Ba$_{0.3}$Fe$_{0.9}$Ti$_{0.1}$O$_3$  bulk materials which have been de-convoluted into two peaks at around 529 eV and 531 eV. The lower binding energy peak is associated with the intrinsic O 1s core spectra and higher energy peak is ascribed to the oxygen vacant sites of this ceramic materials \cite{ref23}. Figures \ref{fig7} (b)  demonstrates the O 1s XPS spectra of corresponding Bi$_{0.7}$Ba$_{0.3}$Fe$_{0.9}$Ti$_{0.1}$O$_3$ nanoparticles prepared by ultrasonication. In this case, a symmetrically single XPS peak is observed which indicates the absence of oxygen vacancy in ultrasonically prepared Bi$_{0.7}$Ba$_{0.3}$Fe$_{0.9}$Ti$_{0.1}$O$_3$ nanoparticles. 
\begin{figure}[!hh]
	\centering
	\includegraphics[width=8.5cm]{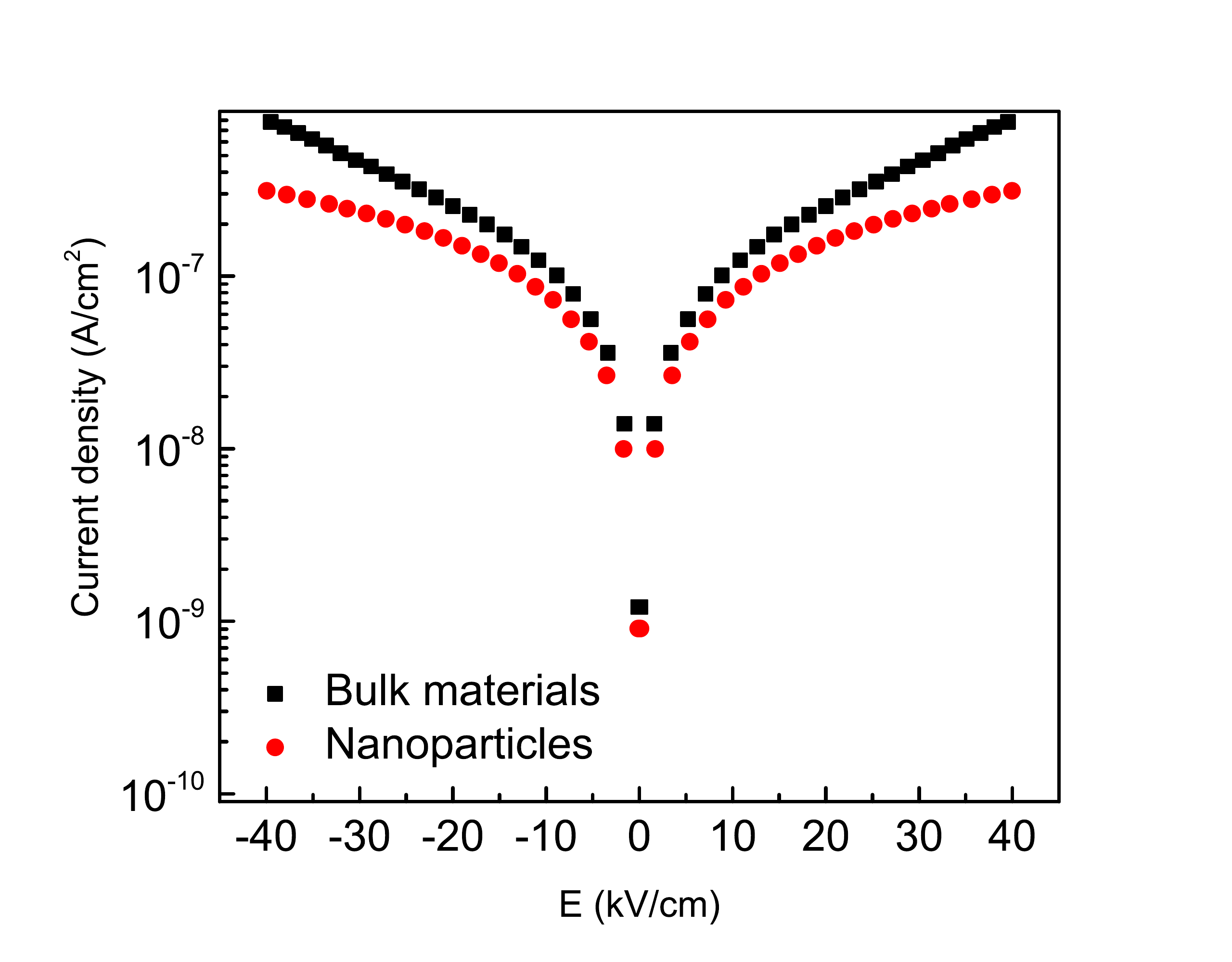}
	\caption{Current density of  Bi$_{0.7}$Ba$_{0.3}$Fe$_{0.9}$Ti$_{0.1}$O$_3$ (a) bulk polycrystalline powder materials and (b) nanoparticles sonicated for 60 minutes.} \label{fig9}
\end{figure}

To compare the leaky behavior of representative Bi$_{0.7}$Ba$_{0.3}$Fe$_{0.9}$Ti$_{0.1}$O$_3$ bulk polycrystalline powder materials and their corresponding nanoparticles, leakage current density, I$_d$ versus electric field, E measurements were performed for an applied field of up to 40 kV/cm. Figure \ref{fig9} shows that the leakage current density of  Bi$_{0.7}$Ba$_{0.3}$Fe$_{0.9}$Ti$_{0.1}$O$_3$ bulk materials is much higher than their corresponding nanoparticles. The high leakage current of bulk materials is predominantly connected with impurity phases and oxygen vacancies \cite{ref20, ref61} of bulk materials as was observed from XRD and XPS measurements, respectively (figure \ref{fig1} and figure \ref{fig7} (a)). The XRD patterns demonstrate the elimination of impurity phases and XPS analysis confirmed the suppression of oxygen related defects in the synthesized nanoparticles. Therefore, the reduction of leakage current in the ultrasonically prepared nanoparticles is reasonable and consistent.  

\begin{figure}[!hh]
	\centering
	\includegraphics[width=8.5cm]{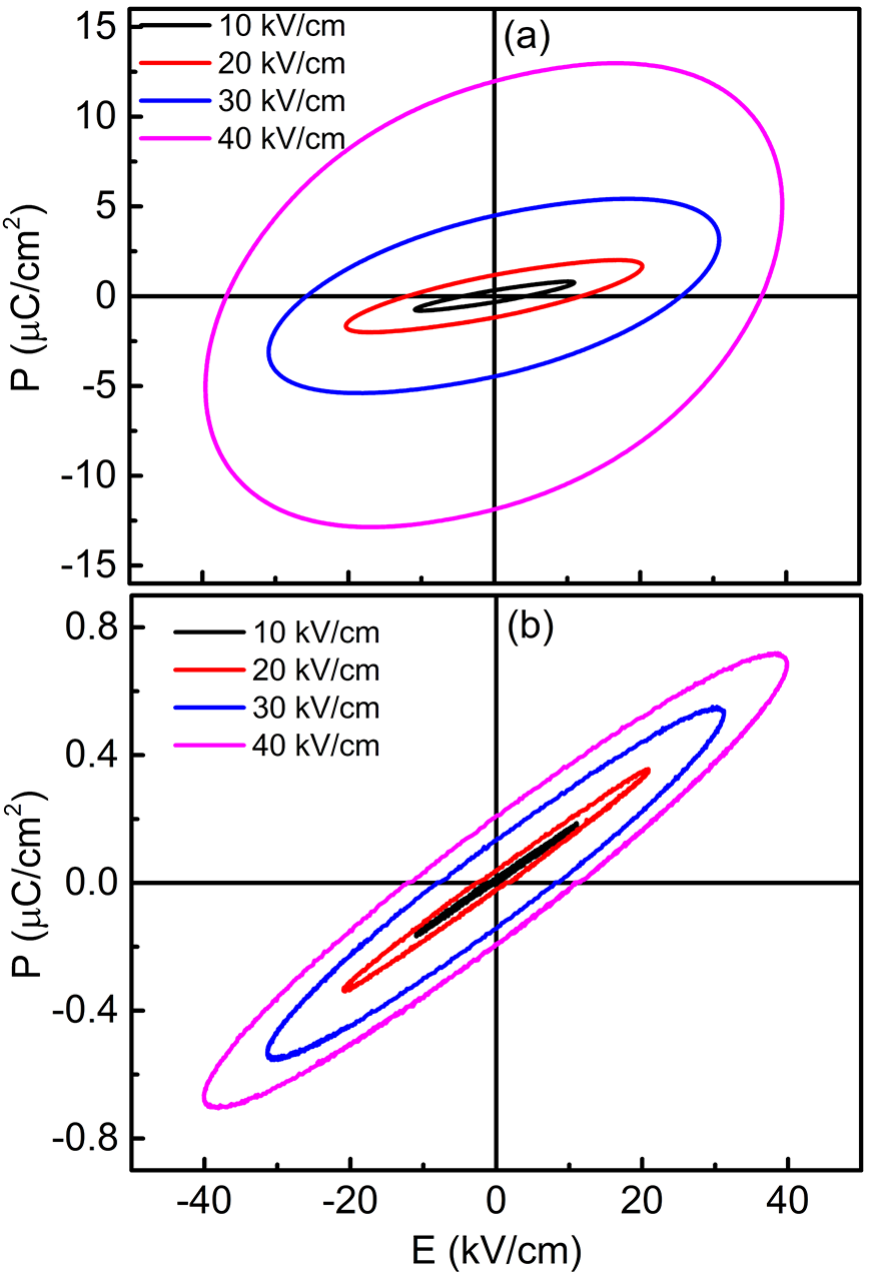}
	\caption{The P-E hysteresis loops of Bi$_{0.7}$Ba$_{0.3}$Fe$_{0.9}$Ti$_{0.1}$O$_3$ (a) bulk polycrystalline powder materials and (b) nanoparticles sonicated for 60 minutes. } \label{fig8}
\end{figure}

We have also carried out ferroelectric measurement to establish polarization versus electric field hysteresis loops (P-E) at applied field with a maximum value of $\pm$ 40 kV/cm for bulk polycrystalline Bi$_{0.7}$Ba$_{0.3}$Fe$_{0.9}$Ti$_{0.1}$O$_3$ powder materials and nanoparticles. The P-E loops of Bi$_{0.7}$Ba$_{0.3}$Fe$_{0.9}$Ti$_{0.1}$O$_3$ bulk materials and their corresponding nanoparticles are shown in figures \ref{fig8} (a) and (b), respectively. The ferroelectricity of bulk Bi$_{0.7}$Ba$_{0.3}$Fe$_{0.9}$Ti$_{0.1}$O$_3$ as well as their corresponding nanoparticles  was evidenced by the P-E loops. The bulk Bi$_{0.7}$Ba$_{0.3}$Fe$_{0.9}$Ti$_{0.1}$O$_3$ materials exhibit a round shaped P-E loop as shown in figure \ref{fig8} (a) due to their high leakage current (figure \ref{fig9}) which is in fact related to oxygen vacancies (figure \ref{fig7} (a)). In the case of Bi$_{0.7}$Ba$_{0.3}$Fe$_{0.9}$Ti$_{0.1}$O$_3$ nanoparticles, figure \ref{fig8} (b), the P-E loops become more and more typical which is expected due to their reduced leakage current density \cite{ref28}. The polarization of the nanoparticles was found to decrease compared to that of bulk materials and this is consistent with previous investigation which reported the reduction of the polarization of BFO with reducing particle size \cite{ref102}.

The P-E loops of bulk Bi$_{0.7}$Ba$_{0.3}$Fe$_{0.9}$Ti$_{0.1}$O$_3$ materials and their corresponding nanoparticles were carried out by varying electric fields. For a fixed driven frequency of 50 Hz, the remanent polarizations of bulk materials as well as ultrasonically prepared nanoparticles were found to increase gradually with an increase in electric field. This is due to the fact that larger electric field provides higher level of driving power responsible for reversal of ferroelectric domains \cite{ref105}.

\section{Conclusions} \label{II}

We have observed a significant difference between the magnetic properties of the bulk powder multiferroic materials and their corresponding nanoparticles. We expect that the observed ferromagnetism in the bulk materials was induced significantly by the presence of magnetic impurity phases. These undesirable impurity phases were completely eliminated in the ultrasonically prepared nanoparticles. The trend of the magnetic behavior of ultrasonically prepared Ba and Ti co-doped BiFeO$_3$ nanoparticles are closely similar to that of undoped bulk BiFeO$_3$ ceramics. The leaky behavior of these bulk powder materials was also suppressed due to ultrasonication for 60 minutes. The suppression of leakage current density improved the ferroelectric behavior of the synthesized nanoparticles. Thus, the present investigation demonstrates the role of impurity phases and high leakage current on the magnetic and ferroelectric behavior of the technologically important multiferroic materials. We believe that the observed improvement of samples purity via the ultrasonic method may be promising to synthesize a wide range of functional materials at nanoscale.

\section{Acknowledgements}
This work was supported by JSPS KAKENHI (Grant No. 26810117) and also by the world Academy of Sciences (TWAS), Ref.:14-066 RG/PHYS/AS-I; UNESCO FR: 324028567. A part of this work was conducted in The Institute for Molecular Science (IMS), supported by Nanotechnology Platform Program (Molecule and Material Synthesis) of the Ministry of Education, Culture, Sports, Science and Technology (MEXT), Japan. The author thank Mr. Tsubasa Sato for TEM imaging.

\end{document}